\documentclass[11pt, a4paper]{article}

\usepackage[utf8]{inputenc}
\usepackage{geometry}
\usepackage{amsmath, amssymb, amsfonts}
\usepackage{graphicx}
\usepackage{hyperref}
\usepackage{natbib}
\usepackage{xcolor}
\usepackage{booktabs}
\usepackage{setspace}
\usepackage{enumitem}
\usepackage{multirow}
\title{\textbf{The Vibe-Check Protocol: Quantifying Cognitive Offloading in AI Programming}}

\author{
Aizierjiang Aiersilan \\
The George Washington University \\
alexandera@gwu.edu \\
}
\date{}

\begin{document}

\maketitle

\begin{abstract}
The integration of Large Language Models (LLMs) into software engineering education has driven the emergence of ``Vibe Coding,'' a paradigm where developers articulate high-level intent through natural language and delegate implementation to AI agents. While proponents argue this approach modernizes pedagogy by emphasizing conceptual design over syntactic memorization, accumulating empirical evidence raises concerns regarding skill retention and deep conceptual understanding. This paper proposes a theoretical framework to investigate the research question: \textit{Is Vibe Coding a better way to learn software engineering?} We posit a divergence in student outcomes between those leveraging AI for acceleration versus those using it for cognitive offloading. To evaluate these educational trade-offs, we propose the \textbf{Vibe-Check Protocol (VCP)}, a systematic benchmarking framework incorporating three quantitative metrics: the \textit{Cold Start Refactor} ($M_{CSR}$) for modeling skill decay; \textit{Hallucination Trap Detection} ($M_{HT}$) based on signal detection theory to evaluate error identification; and the \textit{Explainability Gap} ($E_{gap}$) for quantifying the divergence between code complexity and conceptual comprehension. Through controlled comparisons, VCP aims to provide a quantitative basis for educators to determine the optimal pedagogical boundary: identifying contexts where Vibe Coding fosters genuine mastery and contexts where it introduces hidden technical debt and superficial competence.
\end{abstract}

\section{Introduction}\label{sec:intro}

In the rapidly evolving landscape of computer science education, the advent of Large Language Models (LLMs) has initiated a significant shift in how students engage with programming tasks and conceptual learning. This transformation is most prominently visible in the widespread adoption of a workflow that Andrej Karpathy has termed ``Vibe Coding," a pedagogical and practical approach that involves programmers relinquishing direct syntactic control over code generation and instead focusing their efforts on articulating high-level intent through natural language interactions that guide sophisticated AI agents such as Cursor, Claude Code, and OpenHands \citep{karpathy2025, wang2024openhands}. This shift represents more than a simple tool adoption; it alters the relationship between programmer and code, potentially transforming programming from a direct manipulation activity to a collaborative dialogue between human intent and machine implementation.

This emergent phenomenon prompts a critical pedagogical inquiry: \textit{Is Vibe Coding a superior methodology for learning software engineering principles and practices, or does it merely create a sophisticated simulation of competence that masks fundamental gaps in understanding?} The distinction between these possibilities is subtle yet carries substantial implications for educational policy, curriculum design, and long-term student outcomes. Empirical observations suggest a complex divergence in student learning trajectories, with some learners utilizing these advanced tools as genuine ``force multipliers" that accelerate the implementation of complex architectural patterns. However, a concerning proportion of students seem to fall into patterns characterized by extensive ``cognitive offloading," where they successfully produce functional applications while demonstrating a troubling inability to explain, modify, or extend the underlying logic when AI assistance is removed.

The academic community's response to this dichotomy has been varied. Progressive institutions such as Stanford University have embraced this technological philosophy through innovative courses like CS146S: The Modern Software Developer, which argues that traditional syntax memorization represents an obsolete competency \citep{stanford2025}. However, industry discourse reveals growing concern about the distinction between casual ``Vibe Coding" characterized by rapid, intuition-driven prompting without deep review, and disciplined ``AI-assisted engineering" that maintains rigorous human oversight \citep{shiftmag2025}. Recent benchmarking efforts by \citet{zhao2025vibe} have revealed that while AI agents achieve high functional correctness rates, a significant portion of solutions contain latent defects and security vulnerabilities, raising questions about the long-term viability of code generated through purely AI-driven processes.

From an educational perspective, the risk is the ``illusion of competence'' \citep{rojas2025new}. Rooted in metacognitive biases analogous to the Dunning-Kruger effect \citep{kruger1999unskilled}, this illusion occurs when students mistake the ability to generate code for the ability to understand it. To address this, this paper introduces the \textbf{Vibe-Check Protocol (VCP)}, a novel benchmarking framework designed to quantify the educational impact of Vibe Coding. Unlike prior qualitative studies, VCP employs systematic mathematical modeling to measure skill retention, error detection sensitivity, and conceptual comprehension.

The contribution of this work is twofold. First, it provides a formalized metric set for educators to benchmark Vibe Coding against traditional learning methods. Second, it offers a decision-theoretic framework for curriculum design, helping educators identify the specific pedagogical contexts where AI acceleration is beneficial versus those where traditional syntactic struggle remains essential for deep learning.

The remainder of this paper proceeds as follows: Section \ref{sec:lit_review} reviews the relevant literature. Section \ref{sec:methodology} details the proposed VCP methodology and introduces precise mathematical formulations for skill retention and comprehension. Section \ref{sec:implications} discusses the theoretical implications and expected utility of the framework. Section \ref{sec:limitations} states limitations, and Section \ref{sec:conclusion} concludes.

\section{Literature Review and Problem Statement}\label{sec:lit_review}

\subsection{The Emergence of Vibe Coding and Its Institutional Adoption}\label{subsec:vibe_coding}

Vibe Coding emerges as a direct response to the widespread availability and increasing sophistication of artificial intelligence tools in software development contexts, representing a fundamental reconceptualization of the programmer's role in the code creation process. Karpathy's foundational conceptualization frames this approach as an inherently conversational workflow where users ``give in to the vibes" by expressing their intentions, requirements, and constraints in natural language, thereby allowing Large Language Models to handle the intricate implementation details, syntactic requirements, and low-level optimization decisions that have traditionally consumed substantial developer time and cognitive resources \citep{karpathy2025}. This paradigm represents a significant alignment with broader trends in human-AI collaboration across multiple domains, where tools such as GitHub Copilot, Cursor, and Claude Code enable rapid prototyping and iterative development through sophisticated prompt engineering rather than traditional line-by-line coding approaches that require direct manipulation of syntax and control structures.

From a historical perspective, Vibe Coding builds upon earlier generations of AI-assisted programming tools, including code completion systems, intelligent syntax highlighters, and automated refactoring utilities, but represents a qualitatively different and more fundamental structural change in the programmer's primary role and cognitive engagement with code. Where previous tools like GitHub Copilot focus primarily on suggesting individual functions or completing partial implementations based on immediate context, Vibe Coding involves directing AI agents to construct entire modules, implement complete feature sets, and even architect system-level solutions through iterative dialogue that progressively refines requirements and implementation approaches. This distinction fundamentally reduces the need for direct code manipulation and syntactic expertise, transforming programming from a primarily technical skill into a more collaborative and communicative process that emphasizes intent articulation over implementation execution.

The distinction between these approaches has been carefully documented by industry practitioners who differentiate between casual ``Vibe Coding" characterized by relatively uncritical acceptance of AI suggestions without deep technical review, and sophisticated ``AI-assisted engineering" where human engineers maintain systematic control through structured design documents, comprehensive code review processes, and rigorous test-driven development methodologies \citep{shiftmag2025}. This professional distinction highlights the critical importance of maintaining human oversight and technical judgment even within AI-augmented development environments, suggesting that the most effective applications of these technologies require substantial domain expertise rather than replacing the need for deep programming knowledge.

Institutional adoption of Vibe Coding methodologies has been perhaps most prominently exemplified by Stanford University's innovative CS146S course, which explicitly positions students as orchestrators and directors of AI agents in building modern software systems rather than traditional implementers focused on syntax mastery and algorithmic implementation \citep{stanford2025}. The course curriculum emphasizes advanced prompt crafting techniques, systematic design thinking, iterative refinement processes, and collaborative human-AI workflows, arguing persuasively that such skills represent more relevant and valuable competencies for contemporary industry demands than traditional rote syntax memorization or low-level debugging capabilities. Proponents of this approach contend that Vibe Coding significantly accelerates learning by dramatically lowering entry barriers to complex programming tasks and enabling students to focus their cognitive resources on higher-order thinking, system architecture, and creative problem-solving rather than becoming bogged down in implementation minutiae. Anecdotal reports from early adopters suggest that students have successfully developed sophisticated full-stack applications, complex distributed systems, and innovative software solutions in weeks or months that might traditionally require entire semesters of conventional instruction and practice. However, this remarkable efficiency in project completion potentially circumvents the deliberate practice and productive struggle that expertise research has consistently identified as essential for developing genuine domain mastery, raising fundamental questions about the long-term educational consequences of such accelerated learning approaches \citep{ericsson1993role}.

\subsection{Cognitive Risks and the Illusion of Competence}\label{subsec:cognitive_risks}

Despite efficiency gains, empirical evidence reveals significant pedagogical risks. The concept of ``cognitive offloading'' (where reliance on external aids diminishes intrinsic processing) is well-documented in cognitive psychology \citep{risko2016cognitive}. In educational contexts, this manifests as ``shallow learning,'' where learners achieve surface-level proficiency without the conceptual depth characteristic of genuine expertise \citep{marton1976qualitative}.

Lau and Guo's research on instructor adaptations highlights the risk of students bypassing conceptual understanding \citep{lau2023ban}. This finding is corroborated by Rojas-Galeano's empirical classroom study, which employed a two-phase protocol where 20 undergraduate students first completed programming tasks with AI assistance and then attempted extensions without support \citep{rojas2025new}. Results revealed a notable disconnect: students who rated their understanding as high during the AI-assisted phase showed significantly reduced performance during independent work. Quantitative analysis revealed a moderate-to-strong correlation ($\rho = 0.58$, $p = .008$) between students' self-reported ability to work without AI and their perceived difficulty of the unaided task, suggesting that those who struggled most also recognized their dependency. Qualitative analysis identified common themes including ``partial understanding masked by AI'' ($n=3$) and ``difficulty transferring knowledge,'' with one student noting, ``You think you understood how the code worked, but trying to create it on your own shows you didn't fully get it.''

The ``illusion of competence'' is a cornerstone of this critique. Rooted in the Dunning-Kruger effect, where low-ability individuals overestimate their competence due to lack of metacognitive awareness \citep{kruger1999unskilled}, this illusion manifests when students mistake functional code for mastery. Rojas-Galeano's data showed that while students reported moderate-to-high confidence during AI-assisted tasks, their performance dropped significantly during unaided work, with many struggling to complete simpler extensions of code they had ostensibly created \citep{rojas2025new}.

\subsection{Quality and Reliability Concerns in AI-Generated Code}\label{subsec:quality_concerns}

The implications of Vibe Coding extend beyond educational concerns to production software quality. Zhao et al.'s SUSVIBES benchmark provides a comprehensive evaluation of agent-generated code \citep{zhao2025vibe}. Their findings indicate that while agents like SWE-Agent \citep{yang2024swe} achieve high functional correctness, a significant portion of solutions contain latent defects or vulnerabilities. This suggests that functional tests alone are insufficient to guarantee code quality.

Pearce et al.'s analysis of GitHub Copilot contributions corroborates these findings, demonstrating that AI-generated code can introduce subtle bugs that developers fail to detect \citep{pearce2025asleep}. Industry surveys report that while AI accelerates development, it can create ``trust debt,'' defined as the accumulated burden of code that functions but is not understood \citep{shiftmag2025}. As one CTO observed, AI-generated code ``appears to work perfectly until it catastrophically fails,'' highlighting the necessity of deep comprehension for long-term maintenance.

\subsection{Gaps in Current Evaluation Frameworks}\label{subsec:gaps}

Existing approaches to evaluating AI-generated code and its educational impact suffer from significant limitations. Benchmarks often focus on code quality but neglect educational outcomes entirely. Pedagogical studies like those by Lau and Guo and Rojas-Galeano provide valuable qualitative insights but lack quantitative, scalable metrics for classroom assessment.

Several specific gaps characterize the current landscape. First, there is a lack of longitudinal design, as most studies capture single-session snapshots rather than tracking skill development over time. Second, comparisons between AI-assisted and unassisted performance rarely control for task complexity, leading to an absence of difficulty normalization. Third, educational metrics are often developed ad hoc without establishing reliability or validity, resulting in limited psychometric validation. Fourth, studies identify problems but rarely propose or evaluate corrective strategies, indicating a missing intervention framework. To our knowledge, no framework bridges the gap between code quality assessment and educational outcome measurement.

VCP addresses these gaps by providing a structured protocol that measures learning holistically, incorporates longitudinal tracking, controls for confounding variables, and proposes evidence-based interventions. By integrating error detection metrics (through Hallucination Trap Detection) with skill retention and comprehension measures, VCP aims to provide educators with actionable insights for curriculum design in the AI-assisted learning era.

\section{Methodology: The Vibe-Check Protocol (VCP)}\label{sec:methodology}

\begin{figure}[ht]
    \centering
    \includegraphics[width=1.0\linewidth]{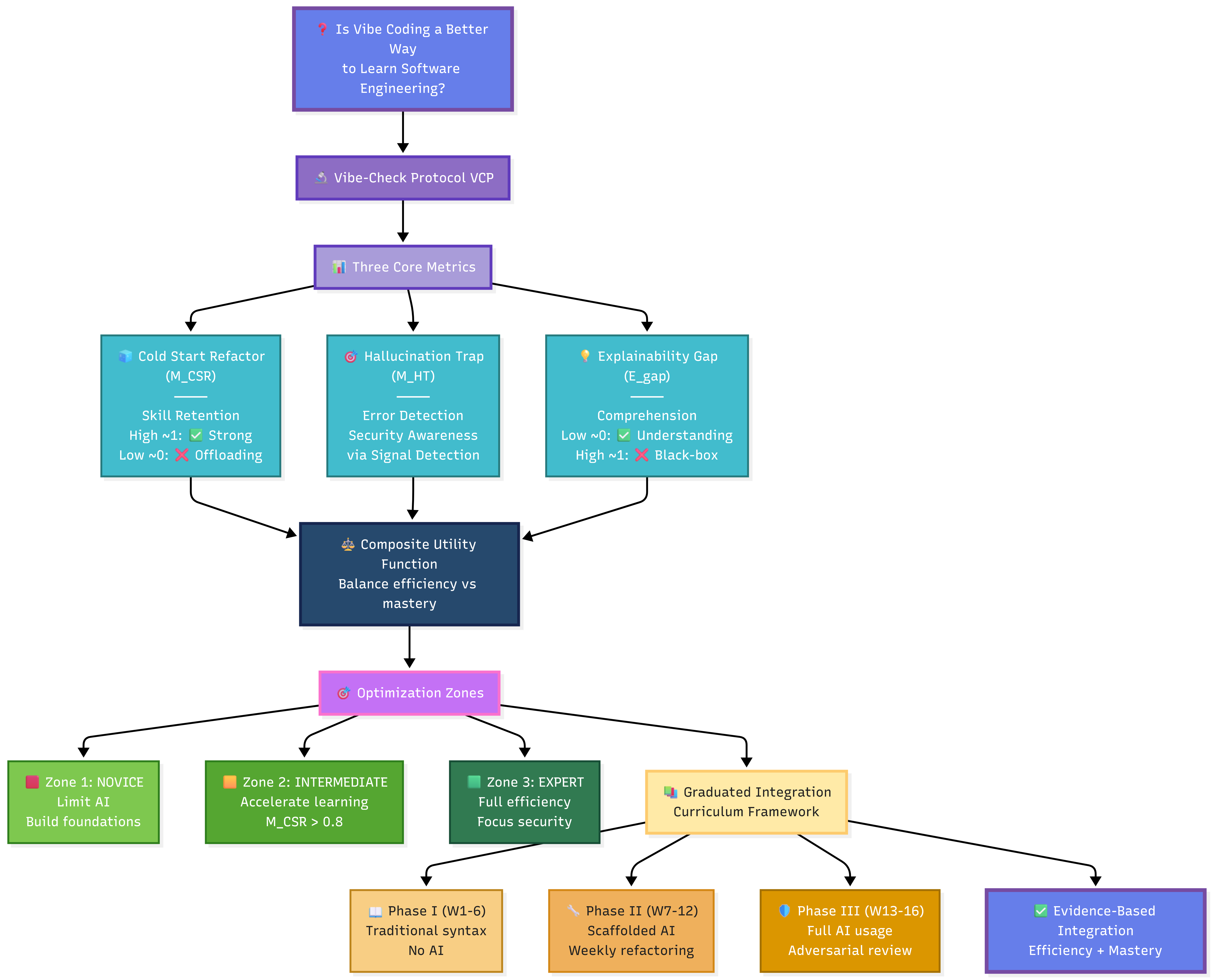}
    \caption{\textbf{The Vibe-Check Protocol (VCP) Framework.} The experimental design compares Traditional Instruction (Control) against Vibe Coding (Experimental) using a mixed-methods approach. The framework quantifies educational impact through three theoretically grounded metrics: (1) \textit{The Cold Start Refactor} ($M_{CSR}$), measuring skill retention via procedural decay; (2) \textit{Hallucination Trap Detection} ($M_{HT}$), assessing vigilance and error sensitivity using Signal Detection Theory; and (3) \textit{The Explainability Gap} ($E_{gap}$), measuring the metacognitive disconnect between generated code complexity and student understanding.}
    \label{fig:framework}
\end{figure}

The Vibe-Check Protocol (VCP) is a framework designed to evaluate the educational impact of Vibe Coding through controlled, longitudinal experiments. Drawing on established practices in educational measurement and insights from recent benchmarks \citep{zhao2025vibe}, VCP comprises three primary metrics targeting distinct dimensions of learning: retention, error detection, and comprehension (see Figure \ref{fig:framework}). The protocol incorporates baseline comparisons with traditionally taught students, controls for task difficulty via standardized rubrics, and employs psychometric validation through pilot testing and inter-rater reliability assessments. Ethical considerations are essential, with all participants providing informed consent and receiving full debriefing on experimental procedures.

\subsection{Theoretical Foundation and Experimental Design}\label{subsec:theoretical_foundation}

The Vibe-Check Protocol is grounded in three complementary theoretical frameworks that collectively address the cognitive, developmental, and metacognitive dimensions of programming education. Cognitive Load Theory \citep{sweller1988cognitive} provides the foundational understanding by distinguishing between intrinsic cognitive load, which represents the inherent complexity of the task itself, extraneous cognitive load resulting from poor instructional design, and germane cognitive load that facilitates the construction of cognitive schemas essential for deep learning. Within this theoretical context, VCP metrics are specifically designed to assess whether AI assistance inadvertently reduces germane cognitive load, thereby potentially undermining the deep learning processes that occur when students engage in productive struggle with challenging programming concepts. This concern is particularly relevant given that the apparent ease of AI-assisted coding may mask the absence of the cognitive processing necessary for skill acquisition.

Complementing this cognitive perspective, Expertise Development Research \citep{ericsson1993role} emphasizes that expert-level performance in any domain requires extensive deliberate practice characterized by immediate feedback, progressive difficulty, and sustained effort over time. The VCP framework operationalizes these principles by evaluating whether Vibe Coding environments provide the ``desirable difficulties" that are essential for skill acquisition, or whether they circumvent these challenges entirely by automating precisely those aspects of programming that would otherwise promote meaningful learning and skill development. This evaluation is critical because the efficiency gains from AI assistance may come at the cost of the procedural knowledge and problem-solving skills that define programming competency.

The third theoretical pillar draws from research on Metacognitive Monitoring \citep{kruger1999unskilled}, which examines the often problematic calibration between perceived competence and actual ability. This body of research demonstrates that individuals with lower ability in a domain often exhibit the greatest overconfidence in their capabilities, a phenomenon directly relevant to AI-assisted learning environments where functional code generation may create an illusion of understanding. The VCP framework addresses this challenge through the Explainability Gap metric, which provides a quantitative comparison between self-assessed understanding and demonstrated comprehension of code functionality and design principles. This comprehensive theoretical integration ensures that VCP captures multiple critical dimensions of the learning process rather than relying on single outcome measures that may miss important aspects of educational impact.

The experimental implementation of VCP employs a sophisticated mixed-methods approach that combines rigorous quantitative metrics with in-depth qualitative interviews, building upon the validated two-phase protocol established by Rojas-Galeano \citep{rojas2025new}. Participants in the proposed studies would be systematically assigned to experimental conditions using Vibe Coding methodologies and traditional control groups employing conventional syntax-focused instruction, with careful matching based on prior programming experience through stratified random assignment procedures. The longitudinal design spans multiple instructional sessions across a full academic semester, enabling comprehensive analysis of skill development trajectories and the identification of both immediate and delayed effects of different pedagogical approaches.

Sample size determination adheres to established power analysis procedures commonly employed in educational research, with calculations targeting 80\% statistical power to detect medium effect sizes (Cohen's $d = 0.5$) at a significance level of $\alpha = 0.05$. These parameters yield minimum sample size requirements of approximately 64 participants per group for between-subjects comparisons and 34 participants for within-subjects repeated measures designs. However, accounting for anticipated attrition rates of approximately 20\% based on longitudinal educational studies, the recommended recruitment target is 100 participants per experimental condition to ensure adequate statistical power throughout the study duration.

Task calibration employs a comprehensive multi-dimensional approach to ensure valid difficulty assessment and meaningful cross-condition comparisons. Cyclomatic Complexity serves as the primary quantitative measure of control flow complexity, providing objective assessment of task difficulty independent of subjective factors. Expert panel ratings complement this approach through structured evaluation using a validated three-point scale ranging from basic to advanced difficulty levels, with inter-rater agreement coefficients maintained above $\kappa > 0.75$ to ensure reliability. Additionally, prerequisite concept enumeration systematically identifies required knowledge elements and maps them to specific course learning objectives, ensuring that task complexity aligns appropriately with curricular expectations and student preparation levels. AI tools are standardized across all experimental conditions, with platforms such as Claude 4 Sonnet accessed through the Cursor interface, and comprehensive logging of all human-AI interactions to enable detailed post-hoc analysis of usage patterns, prompt effectiveness, and iterative refinement strategies.

\subsection{Metric 1: The Cold Start Refactor (\texorpdfstring{$M_{CSR}$}{M\_CSR})}\label{subsec:metric_csr}

The Cold Start Refactor assesses skill retention by modeling the decay of procedural knowledge when AI scaffolding is removed. We posit that skill retention follows an exponential decay function, analogous to Ebbinghaus's forgetting curve \citep{ebbinghaus1913}, and consistent with Cognitive Load Theory's emphasis on schema acquisition \citep{sweller1988cognitive}.

\subsubsection{Mathematical Formulation}
Let $S(t)$ represent the retained skill level at time $t$ after the initial AI-assisted implementation. We model this as:
\begin{equation}
    S(t) = S_0 \cdot e^{-\lambda t}
\end{equation}
where $S_0$ is the initial functional proficiency (typically 1.0 for working code) and $\lambda$ is the decay constant.

In the VCP protocol, we measure the \textit{Reconstruction Velocity} ($V_{rec}$) during an unassisted refactor task at time $t=\Delta t$ (e.g., 24 hours). The metric $M_{CSR}$ is defined as the ratio of unassisted reconstruction velocity to the initial AI-assisted build velocity, scaled by task complexity to compensate for the inherent speed advantage of AI generation in complex tasks:
\begin{equation}
    M_{CSR} = \frac{V_{rec}}{V_{build}} \cdot \Omega(C)
\end{equation}
where $\Omega(C)$ is a complexity weighting function derived from the Cyclomatic Complexity ($CC$) and Halstead Volume ($V$) of the code:
\begin{equation}
    \Omega(C) = \alpha \cdot \ln(CC) + \beta \cdot V
\end{equation}
Here, $\alpha$ and $\beta$ are coefficients derived from baseline data of expert manual developers. A value of $M_{CSR} \approx 1$ implies that the student has internalized the logic sufficiently to reproduce it with expert-level proficiency, indicating high retention. Values approaching 0 indicate severe cognitive offloading.

\subsection{Metric 2: Hallucination Trap Detection (\texorpdfstring{$M_{HT}$}{M\_HT})}\label{subsec:metric_ht}

Hallucination Trap Detection evaluates the student's vigilance and code review capability. We model this using Signal Detection Theory (SDT) \citep{green1966signal} to distinguish between genuine sensitivity to errors and mere response bias (e.g., a student who blindly accepts or blindly rejects code).

\subsubsection{Mathematical Formulation}
We define the student's sensitivity index, $d'$, as the distance between the signal (error) and noise (correct code) distributions:
\begin{equation}
    d' = Z(\text{Hit Rate}) - Z(\text{False Alarm Rate})
\end{equation}
where $Z(\cdot)$ is the inverse cumulative distribution function of the standard normal distribution.
\begin{itemize}
    \item \textbf{Hit Rate ($H$)}: The probability of correctly identifying an injected vulnerability (e.g., SQL injection, race condition).
    \item \textbf{False Alarm Rate ($F$)}: The probability of flagging correct code as erroneous.
\end{itemize}

The metric $M_{HT}$ is normalized to a $[0, 1]$ scale using a sigmoid function centered at a professional competency threshold $\delta$:
\begin{equation}
    M_{HT} = \frac{1}{1 + e^{-k(d' - \delta)}}
\end{equation}
where $k$ determines the steepness of the transition. This formulation penalizes students who achieve high detection rates simply by guessing (which increases False Alarms) and rewards precise discrimination of subtle logic flaws.

\subsection{Metric 3: The Explainability Gap (\texorpdfstring{$E_{gap}$}{E\_gap})}\label{subsec:metric_egap}

The Explainability Gap quantifies the disconnect between the complexity of the generated code and the student's conceptual understanding. We employ an Information Theoretic approach \citep{shannon1948mathematical}, comparing the entropy of the code to the entropy of the student's explanation.

\subsubsection{Mathematical Formulation}
Let $H(C)$ be the Shannon entropy of the code's control flow graph, representing the information content required to fully specify its behavior. Let $H(E)$ be the semantic entropy of the student's explanation, calculated via concept mapping against a ground-truth ontology.

The Explainability Gap is defined as:
\begin{equation}
    E_{gap} = 1 - \frac{H(E)}{H(C) + \epsilon}
\end{equation}
where $\epsilon$ is a small regularization term.
\begin{itemize}
    \item If $H(E) \approx H(C)$, then $E_{gap} \to 0$, indicating the student's mental model matches the code complexity.
    \item If $H(E) \ll H(C)$, then $E_{gap} \to 1$, indicating ``black box'' usage where the code contains logic the student cannot articulate.
\end{itemize}

\subsection{Composite Scoring and Optimization}\label{subsec:composite_scoring}

To answer the research question ``Is Vibe Coding Better?'', we define a pedagogical utility function $U$:
\begin{equation}
    U = w_1 \cdot M_{CSR} + w_2 \cdot M_{HT} + w_3 \cdot (1 - E_{gap}) - \gamma \cdot T_{dev}
\end{equation}
where $T_{dev}$ is the development time (the primary benefit of Vibe Coding). This equation allows us to calculate the \textit{Break-Even Point}: the threshold where the efficiency gain ($\gamma \cdot T_{dev}$) justifies the potential loss in retention or security awareness.

Data analysis employs a mixed-effects model:
\begin{equation}
    Y_{ij} = \beta_0 + \beta_1 \cdot \text{Condition}_{ij} + \beta_2 \cdot \text{Time}_{ij} + u_j + \epsilon_{ij}
\end{equation}
allowing us to isolate the specific effect of the Vibe Coding intervention ($\beta_1$) while controlling for individual student variability ($u_j$).

\section{Theoretical Implications and Expected Utility}\label{sec:implications}

\subsection{The Optimization Boundary: When to Vibe?}\label{subsec:optimization_boundary}

The primary objective of this work is not merely a critique of Vibe Coding, but a decision-theoretic framework for its optimal deployment. Our mathematical modeling suggests that the utility of Vibe Coding is nonlinear with respect to learner expertise.

We propose the \textbf{Cognitive Load Optimization Boundary}, a theoretical construct defined by the intersection of the Skill Decay curve ($S(t)$) and the Complexity Management curve, which delineates optimal pedagogical strategies across different levels of student expertise. This boundary concept recognizes that the educational value of Vibe Coding varies substantially depending on the learner's current competency level and suggests that uniform application across all skill levels may be suboptimal.

In the \textbf{Foundational Acquisition Zone}, which encompasses novice programmers where initial skill level $S_0$ remains relatively low, the introduction of Vibe Coding methodologies typically results in rapid skill decay approaching $\lambda \to \infty$, effectively representing immediate loss of procedural knowledge. Within this developmental stage, what might initially appear as extraneous cognitive load associated with syntax mastery and low-level implementation details actually serves as germane cognitive load essential for fundamental schema formation and conceptual understanding. The struggle with syntax, debugging, and basic algorithmic thinking represents productive difficulty that builds the foundational mental models necessary for subsequent programming competency. Consequently, Vibe Coding interventions should be strictly limited during this phase to preserve the essential learning processes that occur through direct engagement with code.

The \textbf{Architectural Exploration Zone} emerges when students reach intermediate competency levels, specifically when Cold Start Refactor metrics ($M_{CSR}$) stabilize above 0.8, indicating sufficient retention of fundamental programming concepts. At this developmental stage, Vibe Coding transforms into a powerful pedagogical accelerator that enables students to manipulate higher-order abstractions and engage with increased code complexity ($H(C)$) without becoming overwhelmed by implementation minutiae. Within this zone, the Explainability Gap ($E_{gap}$) serves as the primary control variable and monitoring mechanism. As long as $E_{gap}$ remains below 0.3, indicating that student understanding closely matches code complexity, Vibe Coding methodologies enhance learning outcomes by dramatically expanding the scope and sophistication of achievable programming projects, thereby exposing students to architectural patterns and system-level thinking that would otherwise be inaccessible given time and complexity constraints.

Finally, the \textbf{Professional Efficiency Zone} addresses advanced students approaching expert-level competency, where the primary educational concern shifts fundamentally from basic coding ability to critical evaluation and quality assurance skills. For these learners, Hallucination Trap Detection ($M_{HT}$) becomes the most critical assessment metric, as the primary risk transitions from ``not knowing how to code" to ``not recognizing subtle but significant failures" in AI-generated solutions. This zone emphasizes the development of professional-level code review capabilities, security awareness, and the sophisticated judgment required to effectively collaborate with AI systems while maintaining engineering standards and detecting potential vulnerabilities or logic errors that functional testing might miss.

\subsection{Implications for Pedagogical Practice and Curriculum Design}\label{subsec:pedagogical_implications}

The theoretical application and anticipated empirical validation of the Vibe-Check Protocol generates several significant implications for contemporary pedagogical practice in computer science education, fundamentally challenging the assumption that AI-assisted learning tools provide uniformly positive educational outcomes. The framework's central finding suggests that Vibe Coding represents neither an inherently beneficial nor detrimental educational approach, but rather a pedagogical tool whose value depends critically on the cognitive optimization boundary described earlier. This nuanced understanding requires educators to move beyond simple binary decisions about AI tool adoption toward more sophisticated, context-sensitive integration strategies that account for student developmental stages, learning objectives, and long-term competency goals.

The systematic evaluation capabilities provided by VCP metrics enable evidence-based curriculum design that balances the undeniable efficiency advantages of AI-assisted development against the documented risks of skill decay and superficial understanding. While empirical evidence consistently demonstrates that Vibe Coding dramatically accelerates initial code production and enables students to engage with more sophisticated projects earlier in their educational journey, the framework simultaneously reveals that traditional instructional methods demonstrate superior outcomes for fundamental skill retention as measured by the Cold Start Refactor metric ($M_{CSR}$) and critical error detection capabilities assessed through Hallucination Trap Detection ($M_{HT}$), particularly during early developmental stages when foundational competencies are being established.

The quality and security dimensions documented in recent benchmarking studies such as SUSVIBES \citep{zhao2025vibe} add urgency to these pedagogical considerations, as a significant proportion of functionally correct AI-generated solutions contain subtle but potentially serious latent defects and security vulnerabilities. Students who receive training exclusively through Vibe Coding methodologies without complementary critical evaluation skills may graduate with impressive portfolio projects but lack the sophisticated review capabilities necessary for professional software development contexts. This finding aligns closely with emerging industry consensus that distinguishes between casual ``Vibe Coding" characterized by uncritical acceptance of AI output and professional ``AI-assisted engineering" that maintains rigorous human oversight and quality assurance standards \citep{shiftmag2025}.

\subsection{Recommended Curricular Adaptations}\label{subsec:curricular_adaptations}

Based on the theoretical framework and empirical insights provided by VCP metrics, we propose a \textbf{Graduated Integration Framework} that systematically introduces AI-assisted programming methodologies across three carefully structured developmental phases, each targeting specific learning objectives and competency benchmarks. This framework acknowledges that optimal educational outcomes require strategic timing and scaffolding of AI integration rather than uniform application across all learning contexts.

The initial \textbf{Syntax and Semantics Phase}, spanning the first six weeks of instruction, emphasizes traditional pedagogical approaches with strict limitations on AI assistance to ensure fundamental skill acquisition. During this critical foundation-building period, the primary educational objective centers on minimizing the Explainability Gap ($E_{gap}$) for relatively simple code units, ensuring that students develop comprehensive understanding of basic programming constructs, control structures, and algorithmic thinking patterns. The temporary prohibition of AI tools during this phase serves to eliminate the risk of cognitive offloading that could interfere with essential schema formation, requiring students to engage directly with syntax, debugging processes, and fundamental problem-solving strategies that form the bedrock of programming competency.

The subsequent \textbf{Scaffolded Acceleration Phase}, extending from weeks seven through twelve, introduces Vibe Coding methodologies in carefully controlled contexts, specifically targeting boilerplate code generation and automated testing frameworks while maintaining human oversight and comprehension requirements. The central goal during this intermediate phase focuses on maintaining Cold Start Refactor performance above the critical threshold of $M_{CSR} > 0.8$, indicating sufficient skill retention to support more advanced learning objectives. To ensure continued engagement with core programming concepts, students must complete weekly manual refactoring exercises where they systematically modify, extend, and improve AI-generated code segments, thereby demonstrating understanding while benefiting from accelerated development capabilities.

The culminating \textbf{Critical Review Phase}, occupying weeks thirteen through sixteen, permits full utilization of Vibe Coding methodologies while shifting educational emphasis toward maximizing Hallucination Trap Detection capabilities ($M_{HT}$) and developing professional-level quality assurance skills. During this advanced phase, students assume the role of technical auditors responsible for systematically evaluating AI agent output, specifically focusing on identifying logic errors, security vulnerabilities, and subtle defects of the type documented in comprehensive benchmarks such as SUSVIBES \citep{zhao2025vibe}. This approach transforms students from passive consumers of AI-generated code into active collaborators capable of leveraging AI capabilities while maintaining critical oversight and professional engineering standards.

\subsection{The Professional Distinction: AI-Assisted Engineering versus Vibe Coding}\label{subsec:professional_distinction}

As discussed in Section \ref{subsec:vibe_coding}, industry discourse distinguishes between casual ``Vibe Coding" and disciplined ``AI-assisted engineering." The Vibe-Check Protocol operationalizes this distinction, enabling systematic assessment of whether students are developing the critical thinking and quality assurance capabilities necessary for the latter approach rather than simply becoming efficient consumers of AI-generated output.

High performance on Hallucination Trap Detection metrics ($M_{HT}$) serves as a quantitative indicator that students are functioning as true AI-assisted engineers rather than passive recipients of automated solutions. Such students demonstrate the vigilance, technical judgment, and systematic evaluation skills required to identify subtle logic errors, potential security vulnerabilities, and design flaws that may not be apparent through functional testing alone. Conversely, low Explainability Gap values ($E_{gap}$) provide evidence of genuine conceptual ownership and understanding that extends far beyond surface-level acceptance of working code. Together, these metrics distinguish between students who can effectively collaborate with AI systems while maintaining engineering standards and those who have developed dependence relationships that may prove problematic in professional contexts.

This distinction addresses a critical concern documented in recent empirical research, where student preferences and self-assessments reveal awareness of the importance of maintaining fundamental coding skills even in AI-augmented development environments. As reported by Rojas-Galeano, approximately 60\% of surveyed students explicitly agreed that ``coding skills are needed to adapt AI output" \citep{rojas2025new}, suggesting that students themselves recognize the limitations of purely AI-dependent approaches. The VCP framework provides the systematic assessment infrastructure necessary to ensure that curricula support these student aspirations by transforming the documented ``illusion of competence" into measurable, verifiable mastery that aligns with both educational objectives and professional requirements.

\subsection{Targeted Intervention Strategies and Remediation Approaches}\label{subsec:interventions}

The diagnostic capabilities of the Vibe-Check Protocol enable the development and implementation of targeted educational interventions based on specific student performance profiles across the three core metrics, moving beyond one-size-fits-all approaches toward personalized remediation strategies that address individual learning needs and skill gaps. Students exhibiting high decay rates as measured by the Cold Start Refactor metric would benefit from systematic spaced retrieval practice interventions that incorporate progressively increasing intervals of AI-free coding exercises, thereby strengthening procedural memory consolidation and reducing dependence on external scaffolding. These interventions might include weekly unassisted refactoring challenges, timed coding exercises that gradually increase in complexity, and structured reflection activities that help students internalize the problem-solving strategies initially supported by AI assistance.

For students demonstrating low sensitivity scores in Hallucination Trap Detection, explicit code review training becomes essential, focusing on systematic exposure to documented anti-patterns, common vulnerability types, and subtle defect categories that frequently escape automated detection systems. Such training should emphasize real-world failure modes including timing attacks that exploit race conditions, various injection vulnerabilities affecting database and system security, and authentication bypass techniques that compromise access controls. The curriculum should incorporate case studies from actual security incidents, structured analysis of vulnerable code examples, and hands-on experience with both static and dynamic analysis tools commonly employed in professional development environments.

Beyond remediation, the VCP framework supports identification and development of positive learning trajectories that leverage students' existing strengths while addressing areas for improvement. Preliminary observations suggest that prompt engineering quality improves substantially over time as students develop more precise, context-rich communication strategies with AI systems, indicating metacognitive growth in their understanding of how to effectively collaborate with automated tools. Similarly, architectural sophistication appears to increase as Vibe Coding methodologies enable earlier engagement with system-level design patterns and complex software architectures that would otherwise remain inaccessible due to implementation complexity barriers. These positive developments support a balanced perspective on AI-assisted learning, recognizing that while Vibe Coding introduces documented risks requiring careful management, it also creates unique educational opportunities that can enhance rather than undermine learning outcomes when properly scaffolded and systematically monitored.

\section{Limitations and Future Research}\label{sec:limitations}

\subsection{Current Limitations and Methodological Constraints}\label{subsec:current_limitations}

The Vibe-Check Protocol, while representing a substantial advancement over existing approaches to evaluating AI-assisted programming education, faces several significant limitations that require careful acknowledgment and consideration for future research directions. The most fundamental constraint of this work lies in its current theoretical status, as the VCP framework awaits full empirical validation through controlled classroom experiments. This limitation stems from the complex intersection of institutional review board requirements, the extended timelines necessary for longitudinal educational research, and the rapidly evolving landscape of AI tools that creates moving targets for systematic study. The absence of empirical validation means that the proposed thresholds, weightings, and mathematical formulations require confirmation through real-world implementation before their practical utility can be fully established.

Generalizability constraints present another significant methodological concern, as future pilot studies will necessarily involve students from specific institutional and cultural contexts that may not represent the broader diversity of programming education environments. The applicability of proposed thresholds and metric weightings to diverse populations, including non-traditional students pursuing career transitions, international cohorts with varying educational backgrounds, and professional retraining contexts with different motivational structures, remains to be empirically established. These population differences may significantly influence the validity of VCP metrics and the interpretation of results across different educational settings.

Methodological threats to validity also include potential Hawthorne effects, where the process of observation and measurement may inadvertently alter participant behavior in ways that compromise the authenticity of assessment results. This concern is particularly acute for the Hallucination Trap Detection metric ($M_{HT}$), where student awareness of error detection evaluation could artificially inflate detection rates through increased vigilance that may not persist in natural learning environments. Future implementations must therefore explore sophisticated unobtrusive measurement strategies and control conditions that can mitigate these observational biases while maintaining the integrity of the assessment process.

The rapidly evolving nature of artificial intelligence capabilities introduces another layer of complexity to the validation and application of VCP metrics. AI tools and their capabilities are advancing at rapid rates, meaning that the specific systems and performance characteristics being evaluated today may differ substantially from those available in subsequent academic semesters or years. This technological fluidity necessitates ongoing calibration and validation of VCP metrics as model capabilities advance, creating a continuous requirement for framework adaptation to ensure that measurements remain relevant, accurate, and aligned with current AI tool capabilities and limitations.

Resource intensity represents a significant practical barrier to widespread adoption of the complete VCP framework in educational settings. Full implementation requires substantial investments in trained human raters capable of consistent and reliable assessment, sophisticated longitudinal tracking infrastructure for managing multi-semester data collection, and considerable instructor time for implementation and ongoing monitoring. The scalability challenges become particularly acute when considering application to large enrollment courses or resource-constrained educational institutions. Without the development of additional automated assessment tools or significantly streamlined protocols, the practical utility of VCP may remain limited to research contexts rather than achieving broader pedagogical impact.
And the integration between security awareness and pedagogical assessment remains incomplete within the current framework. While VCP incorporates security considerations through the Hallucination Trap Detection metric, the framework does not yet provide comprehensive bridging to professional security assessment standards and industry vulnerability detection training protocols. This gap represents a missed opportunity to better align educational assessment with professional preparation, particularly given the increasing importance of security awareness in software development careers and the documented presence of vulnerabilities in AI-generated code as identified in recent benchmarking studies.

\subsection{Future Research Directions and Methodological Extensions}\label{subsec:future_research}

The development and validation of the Vibe-Check Protocol opens numerous avenues for future research that address current limitations while expanding the framework's scope and applicability. Cross-institutional validation represents the most immediate and critical research priority, requiring the coordination of multi-site studies involving diverse student populations across different institutional contexts, geographic regions, and cultural backgrounds. Such comprehensive validation efforts are essential for establishing normative performance data and determining the generalizability of proposed thresholds and weightings across varied educational environments. These studies should systematically examine how institutional factors such as class size, instructor experience, technological infrastructure, and pedagogical traditions influence VCP metric performance and interpretation.

The integration of advanced physiological and cognitive assessment measures offers promising opportunities for developing deeper insights into the learning processes underlying AI-assisted programming education. The systematic incorporation of eye-tracking technology, structured think-aloud protocols, and neuroimaging approaches could provide enhanced visibility into cognitive engagement patterns that extend far beyond observable behavioral outcomes. Such multi-modal assessment strategies would enable researchers to distinguish between genuine conceptual comprehension and superficial pattern matching behaviors, potentially revealing cognitive processes that current behavioral measures cannot detect. This enhanced understanding could lead to more nuanced interpretation of VCP metrics and identification of additional assessment dimensions that complement existing measures.

Addressing the scalability challenges identified in current limitations requires substantial investment in automated assessment tool development, leveraging advances in machine learning and natural language processing to reduce the human resource requirements of VCP implementation. Automated approaches could potentially handle Explainability Gap ($E_{gap}$) coding through sophisticated semantic analysis of student explanations, generate appropriate Hallucination Trap Detection scenarios tailored to specific code contexts, and provide real-time feedback mechanisms that support both student learning and instructor decision-making. The development of such tools would transform VCP from a resource-intensive research instrument into a practical classroom assessment system suitable for widespread adoption.

Longitudinal studies that track professional outcomes represent another crucial research direction for establishing the predictive validity and real-world relevance of VCP assessments. Following graduates into industry positions over multiple years would provide essential evidence regarding the relationship between VCP performance and subsequent professional competency, career advancement, and long-term technical contributions. Such studies could validate whether high performance on VCP metrics translates into measurable advantages in professional software development contexts, thereby establishing the framework's utility for both educational assessment and career preparation evaluation.

Intervention optimization through systematic randomized controlled trials offers opportunities to identify the most effective pedagogical strategies for different learner profiles and educational contexts. Comparative studies examining various scaffolding approaches, timing strategies, and instructional modifications could determine optimal implementation parameters for the Graduated Integration Framework across diverse student populations. These investigations should systematically vary factors such as the duration of AI-free instruction, the types of scaffolded activities during transition phases, and the complexity progression of evaluation tasks to identify evidence-based best practices for AI integration in programming education.

Cultural adaptation research represents a critical but currently underexplored dimension of VCP validation, as attitudes toward technology, collaboration norms, educational traditions, and learning preferences vary significantly across international contexts. Understanding how cultural factors moderate VCP findings and influence the interpretation of metrics could reveal important boundary conditions for framework applicability and suggest necessary adaptations for global implementation. Such research might examine how individualistic versus collectivistic cultural orientations influence collaborative AI use, how different educational traditions shape student expectations regarding technological assistance, and how varying attitudes toward authority and expertise affect students' critical evaluation of AI-generated solutions.
Besides, expanding safety training integration within the VCP framework offers opportunities to strengthen the connection between educational assessment and professional preparation in cybersecurity-conscious development environments. Future research should develop comprehensive curriculum modules that systematically bridge VCP security awareness measures with established professional secure coding practices, potentially incorporating repository-level challenges similar to those employed in the SUSVIBES benchmark \citep{zhao2025vibe}. This expanded focus on security could transform VCP from a general programming education assessment tool into a specialized framework for preparing students to work effectively and safely with AI assistance in safety-critical development contexts.

\section{Conclusion}\label{sec:conclusion}

The Vibe-Check Protocol offers a systematic framework for evaluating the educational impact of AI-assisted programming. By quantifying skill retention, error detection, and conceptual understanding, VCP addresses the critical need for evidence-based assessment in the era of Vibe Coding. This framework moves beyond simple functional correctness to capture the cognitive dimensions of learning, distinguishing between genuine mastery and the illusion of competence.

Ultimately, the goal of VCP is to guide the integration of AI in education. By identifying the pedagogical boundaries where AI acceleration supports rather than supplants deep learning, educators can foster a generation of software engineers who leverage AI as a powerful tool while maintaining the fundamental expertise required for robust, secure, and innovative system design.

\newpage
\bibliographystyle{plainnat}
\bibliography{ref}

\end{document}